\documentclass[12pt]{iopart}
\pdfoutput=1
\usepackage{iopams}
\usepackage{graphicx}
\usepackage{color}

\begin{document}

\title{On the critical exponent $\alpha$ of the 5D random--field Ising model}

\author{Nikolaos G. Fytas$^{1}$, V\'{i}ctor Mart\'{i}n-Mayor$^{2,3}$, Giorgio Parisi$^4$, Marco Picco$^5$, and Nicolas Sourlas$^6$}

\address{$^1$Applied Mathematics Research Centre, Coventry
University, Coventry CV1 5FB, United Kingdom}

\address{$^2$Departamento de F\'isica T\'eorica I, Universidad
Complutense, 28040 Madrid, Spain}

\address{$^3$Instituto de Biocomputac\'ion y F\'isica de
Sistemas Complejos (BIFI), 50009 Zaragoza, Spain}

\address{$^4$Dipartimento  di  Fisica,  Sapienza Universit\`{a} di
Roma,  P.le  Aldo  Moro  2, 00185  Rome, Italy and INFN, Sezione
di Roma I,  NANONTEC -- CNR,  P.le  A. Moro 2, 00185 Rome, Italy}

\address{$^5$Laboratoire de Physique Th\'eorique et Hautes Energies, UMR7589, Sorbonne Universit\'e et CNRS, 4 Place Jussieu, 75252 Paris Cedex 05, France}

\address{$^6$Laboratoire de Physique Th\'eorique de l'Ecole
Normale Sup\'erieure (Unit{\'e} Mixte de Recherche du CNRS et de
l'Ecole Normale Sup\'erieure, associ\'ee \`a l'Universit\'e
Pierre et Marie Curie, PARIS VI)
  24 rue Lhomond, 75231 Paris CEDEX 05, France}
  
\ead{nikolaos.fytas@coventry.ac.uk}

\date{\today}

\begin{abstract}
We present a complementary estimation of the critical exponent
$\alpha$ of the specific heat of the 5D random-field Ising model
from zero-temperature numerical simulations. Our result $\alpha =
0.12(2)$ is consistent with the estimation coming from the
modified hyperscaling relation and provides additional evidence in
favor of the recently proposed restoration of dimensional
reduction in the random-field Ising model at $D = 5$.
\end{abstract}

\submitto{Journal of Statistical Mechanics}

\maketitle


The random-field Ising model (RFIM) is one of the archetypal
disordered
systems~\cite{imr75,aharony76,young77,fishman79,parisi79,cardy84,
imbrie84,villain84,bray85,fisher86,schwartz85,gofman93,esser97,barber,young1999},
extensively studied due to its theoretical interest, as well as
its close connection with experiments in condensed-matter
physics~\cite{young1999,by1991,rieger1995,jaccarino,vink}. Its
beauty stems from the combination of random fields and the
standard Ising model that creates rich and complicated physical
phenomena, responsible for a great volume of research over the
last 40 years and more. It is well established that the physically
relevant dimensions of the RFIM lay between $2 < D < 6$, where
$D_{\rm l} = 2$ and $D_{\rm u} = 6$ are the lower and upper
critical dimensions of the model, respectively. For $D \geq D_{\rm
u}$ one expects the standard mean-field
behavior~\cite{imr75,villain84,bray85,fisher86,berker86,bricmont87},
whereas exactly at $D = D_{\rm u}$ the notoriously obscuring
logarithmic corrections appear~\cite{ahrens11}.

In the last few years, the development of a powerful panoply of
simulation and statistical analysis methods~\cite{fytas16} have
set the basis for a fresh revision of the problem. In fact, some
of the main controversies have been resolved, the most notable
being the illustration of critical universality in terms of
different random-field
distributions~\cite{fytas13,fytas16b,fytas17} and the restoration
of supersymmetry and dimensional reduction at $D =
5$~\cite{fytas17b,fytas18,fytas19} (see also
references~\cite{tissier11,tissier12,tarjus13,hikami18} for
additional evidence in this respect).

In particular, the large-scale numerical simulations of the 5D
RFIM reported in reference~\cite{fytas17b} have provided
high-accuracy estimates for the spectrum of critical exponents and
for several universal ratios (see Table III in
reference~\cite{fytas17b}), with one missing element: that of the
direct computation of the critical exponent $\alpha$ of the
specific heat. Let us point out that the specific heat of the RFIM
is of experimental interest~\cite{jaccarino} and that the value of
$\alpha$ has severe implications for the validity of the
fundamental scaling relations, and in particular for the
Rushbrooke relation, $\alpha+2\beta+\gamma = 2$, that has been the
most controversial of
all~\cite{middleton1,hartmann01,theodorakis,fytasEPJB,nowak}.
Therefore a strong command on this aspect of the model's critical
behavior is necessary. In the current work we fill this gap by
performing additional simulations and scaling analysis that allow
us to directly compute $\alpha$ for the 5D RFIM and to therefore
present a complete picture of the scaling behavior of the specific
heat. Our final estimate, $\alpha = 0.12(2)$, agrees well with
that of the 3D Ising universality class,
$0.110087(12)$~\cite{kos16}, and therefore constitutes additional
evidence in favor of our recently proposed restoration of
dimensional reduction at $D = 5$~\cite{fytas17b,fytas18,fytas19}.


The RFIM Hamiltonian is
\begin{equation}
\label{H} {\cal H} = - J \sum_{<xy>} S_x S_y - \sum_{x} h_x S_x \;
,
\end{equation}
with the spins $S_x = \pm 1$ occupying the nodes of a hyper-cubic
lattice in space dimension $D$ with nearest-neighbor ferromagnetic
interactions $J$ and $h_x$ independent random magnetic fields with
zero mean and dispersion $\sigma$. Here we consider the
Hamiltonian~(\ref{H}) on a $D=5$ hyper-cubic lattice with periodic
boundary conditions and energy units $J=1$. Our random fields
$h_{x}$ follow either a Gaussian $({\mathcal P}_G)$, or a
Poissonian $({\mathcal P}_P)$ distribution of the form
\begin{equation}
\label{distribution} {\mathcal P}_G(h,\sigma) = {1\over \sqrt{2
\pi
      \sigma^2}} e^{- {h^2 \over 2\sigma^2}}\;;\ {\mathcal P}_P(h,\sigma) = {1\over 2 |\sigma| } e^{- {|h| \over
\sigma}}\;,
\end{equation}
where $-\infty< h < \infty$ and $\sigma$ the
disorder-strength control parameter.

\begin{figure}
\centerline{\includegraphics[scale=.4, angle=0]{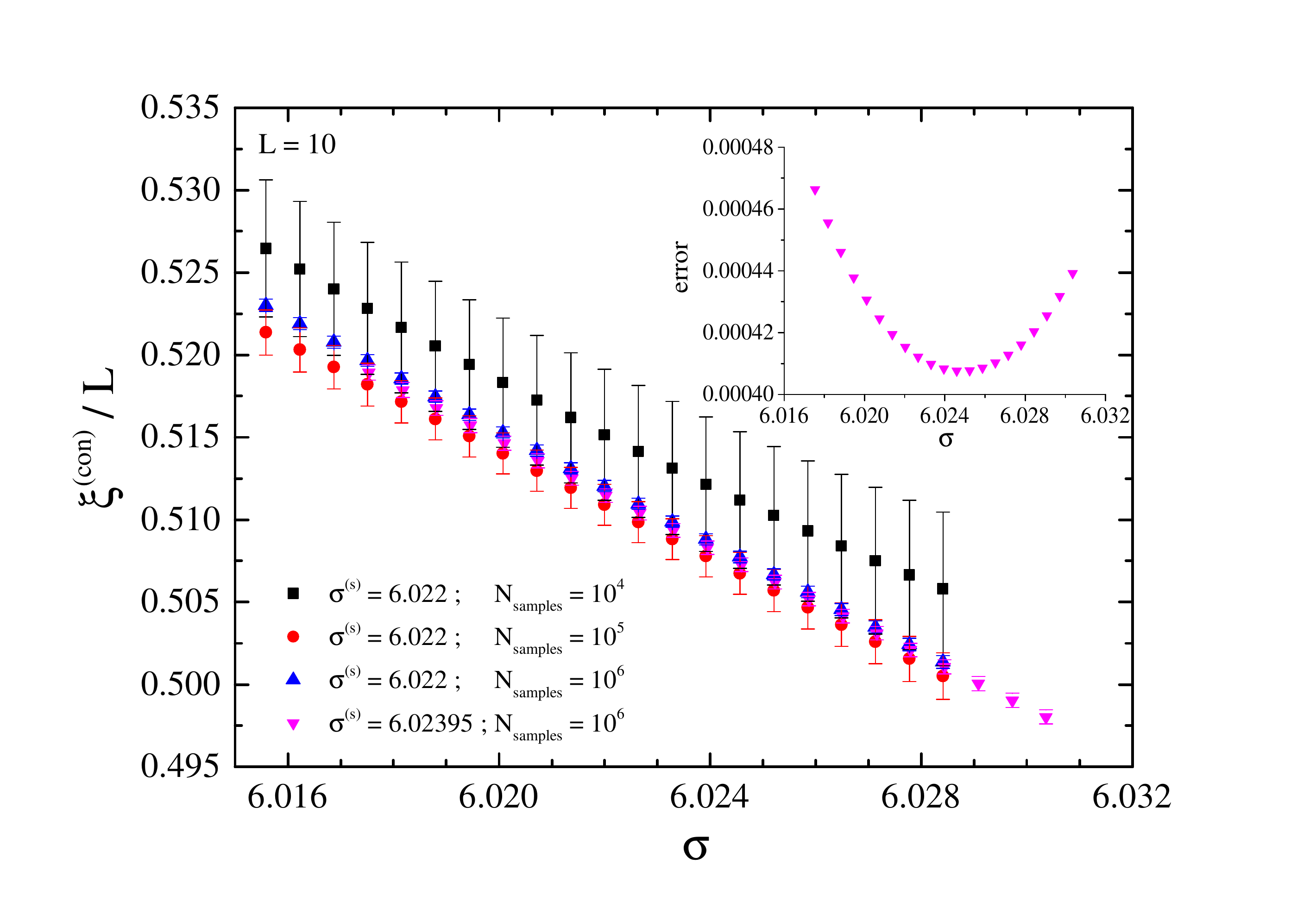}}
\caption{Connected correlation length in units of the system size
$L$ versus $\sigma$ for the 5D Gaussian RFIM and a system of linear size $L = 10$. Four distinct simulation sets are shown, corresponding to different simulation values,
$\sigma^{\rm (s)}$, and different sets of random-field
realizations. The inset illustrates the reweighting error-evolution
for the fourth simulation set with $\sigma = 6.02395$ and $N_{\rm samples} = 10^6$.} \label{fig:1}
\end{figure}

As it is well-established, in order to describe the critical
behavior of the model one needs two correlation functions, namely
the connected and disconnected propagators,
$C^{\mathrm{(con)}}_{xy}$ and $C^{\mathrm{(dis)}}_{xy}$:
\begin{equation}\label{eq:anomalous}
C^{\mathrm{(con)}}_{xy}\!\equiv\!\frac{\partial\overline{\langle
  S_x\rangle}}{\partial h_y} \; ; \,
 C^{\mathrm{(dis)}}_{xy}\!\equiv\!
\overline{\langle S_x\rangle\langle S_y\rangle}\! \,,
\end{equation}
where the $\langle \ldots \rangle$ are thermal mean values as
computed for a given realization, a sample, of the random fields
$\{h_x\}$. Over-line refers to the average over the samples.
Following the prescription of reference~\cite{fytas16}, for each of these two propagators we scrutinize the second-moment correlation lengths, denoted as $\xi^{\mathrm{(con)}}$ and $\xi^{\mathrm{(dis)}}$, respectively.

\begin{figure}
\centerline{\includegraphics[scale=.4, angle=0]{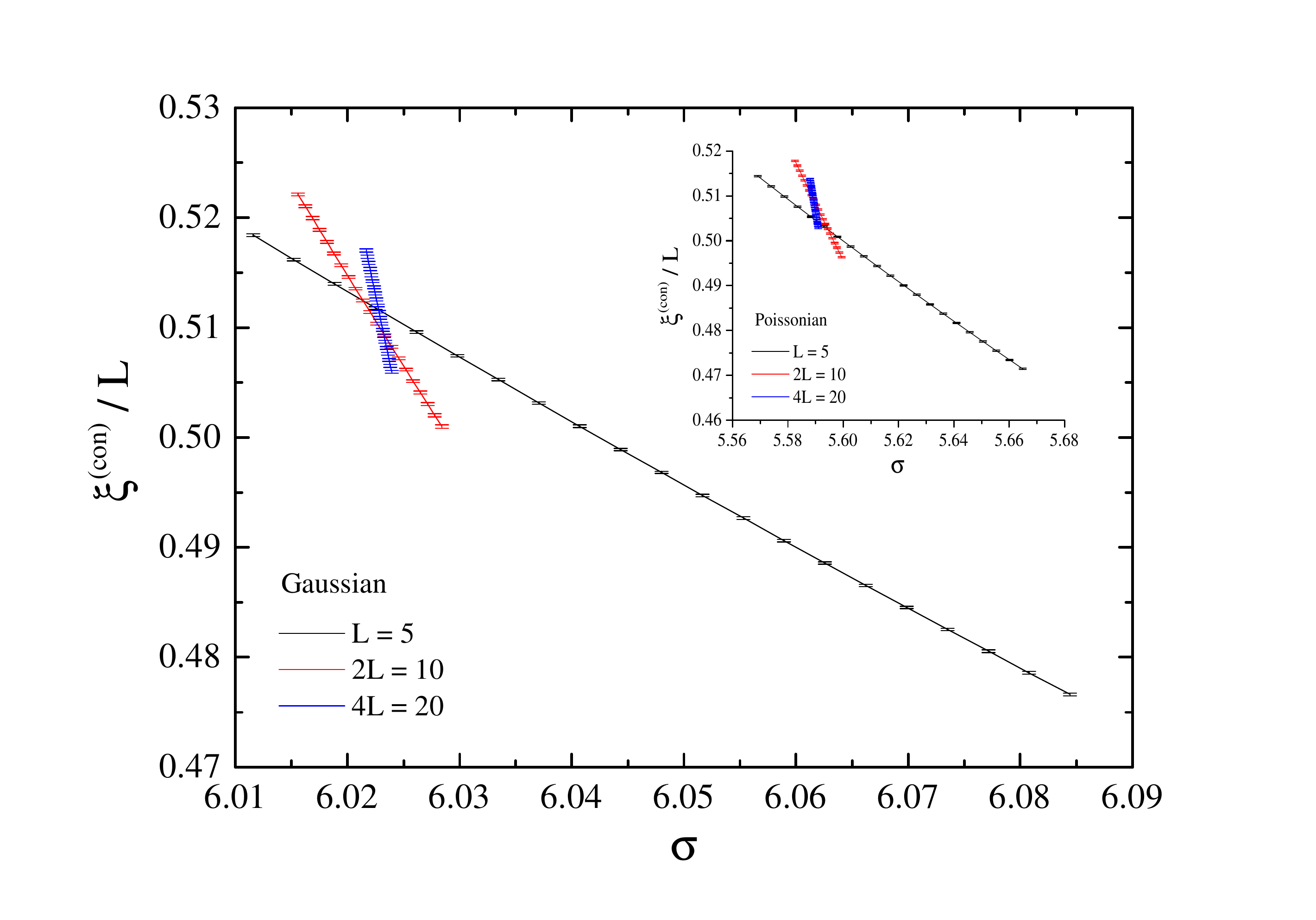}}
\caption{Connected correlation length in units of the system size
$L$ versus $\sigma$ for the 5D Gaussian (main panel) and Poissonian (inset) RFIM. An  illustrative example of the three lattice-size sequence $(L,2L,4L) = (5,10,20)$ used in the application of the modified quotients method is shown [see equations~(\ref{eq:FSS-3L}) and (\ref{eq:QO_new})]. Data taken from reference~\cite{fytas17b}.} \label{fig:2}
\end{figure}

Our numerical simulations for the 5D RFIM are described in
reference~\cite{fytas17b}. We therefore outline here the very
necessary details. We simulated lattice sizes from $L_{\rm min}=4$
to $L_{\rm max}=28$. For each pair of ($L$, $\sigma$) values we
generated ground states for $10^7$ samples -- for the additional
simulations at the most accurate determinations of the critical
points shown below in figures~\ref{fig:4} and \ref{fig:5}, $10^6$
samples were generated -- exceeding previous relevant
studies~\cite{ahrens11} by a factor of $10^3$ on average. The
calculation of the ground states of the RFIM was based on the
well-established mapping~\cite{middleton1,hartmann01,
theodorakis,fytasEPJB,nowak,ogielski85,hartmann95,bastea98,hartmann99,hartmann02,seppala,middleton2,middleton3,
machta03,alava,zumsande,puri11,weigel,fytas12,hartmannbook1,hartmannbook2}
to the maximum-flow problem~\cite{angles,cormen,papadimitriou}. We
used our own C version of the push-relabel algorithm of Tarjan and
Goldberg~\cite{tarjan}, involving some technical modifications
proposed by Middelton and collaborators for further
efficiency~\cite{middleton2,middleton3}. Suitable generalized
fluctuation-dissipation formulas and reweighting extrapolations
have facilitated our analysis, as exemplified in
reference~\cite{fytas16}. A comparative illustration in favor of
the numerical accuracy of our scheme is shown in
figure~\ref{fig:1} for the universal ratio $\xi^{\rm (con)} / L$
of an L = 10 Gaussian RFIM and four different simulation sets, as
outlined in the panel.


The specific heat of the RFIM can be estimated using ground-state
calculations in two complementary frameworks, both based on the
analysis of singularities of the bond-energy density
$E_{J}$~\cite{holm97}. This bond-energy density is the first
derivative $\partial E/\partial J$ of the ground-state energy with
respect to the random-field strength
$\sigma$~\cite{middleton1,hartmann01}. The derivative of the
sample averaged quantity $\overline{E}_{J}$ with respect to
$\sigma$ then gives the second derivative with respect to $\sigma$
of the total energy and thus the sample-averaged specific heat
$C$. The singularities in $C$ can also be studied by computing the
singular part of $\overline{E}_{J}$, as $\overline{E}_{J}$ is just
the integral of $C$ with respect to $\sigma$. Thus, one may
estimate $\alpha$ by studying the behavior of $\overline{E}_{J}$
at $\sigma = \sigma_{\rm c}$~\cite{middleton1} , via the scaling
form
\begin{equation}
\label{eq:E_J_scaling} \overline{E}_{J}(L,\sigma_{\rm c}) = E_{J,\infty} + b
L^{(\alpha-1)/\nu}(1+b'L^{-\omega}),
\end{equation}
where $E_{J,\infty}$, $b$, and $b'$ are non-universal constants,
and $\omega$ is the universal corrections-to-scaling exponent.

\begin{figure}
\centerline{\includegraphics[scale=.4, angle=0]{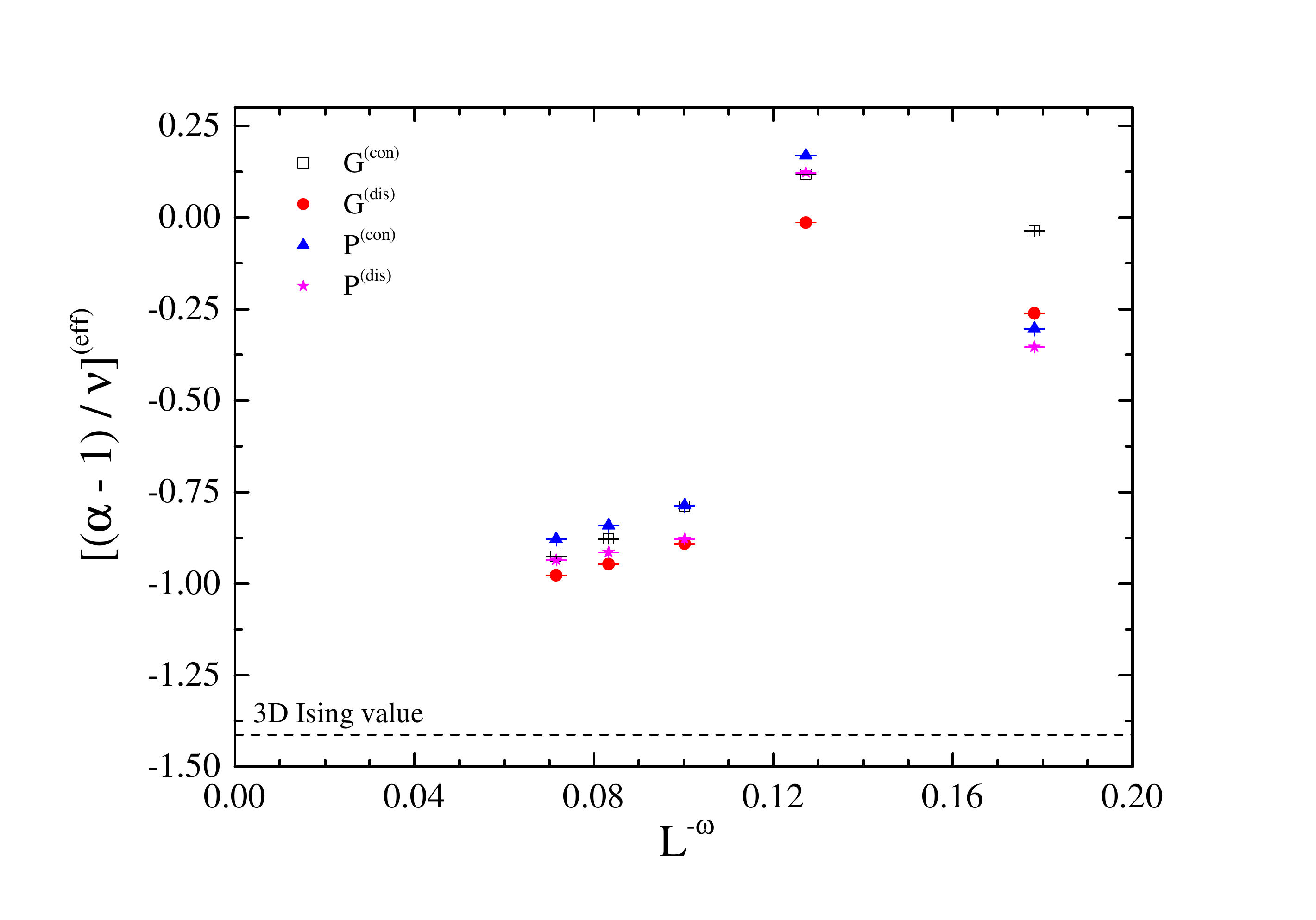}}
\caption{Effective exponent ratio $(\alpha - 1)/\nu$ versus
$L^{-\omega}$ for all random-field distributions and crossing
points considered in this work. Note the notation $\rm {Z}^{\rm
(x)}$, where Z stands for the distribution -- G for Gaussian and P
for Poissonian -- and the superscript x for the connected (con) or
disconnected (dis) type of the universal ratio $\xi^{\rm (x)} /
L$, used for the application of the quotients method [see
equations~(\ref{eq:FSS-3L}) and (\ref{eq:QO_new})].} \label{fig:3}
\end{figure}

Of course, the use of equation~(\ref{eq:E_J_scaling}) for the
application of standard finite-size scaling methods requires an
\emph{a priori} knowledge of the \emph{exact} value of the
critical random-field strength $\sigma_{\rm c}$ [see also the
analysis below in figures~\ref{fig:4} and \ref{fig:5}]. Although
we currently have at hand such high-accuracy estimates of the
critical fields for both types of the random-field distributions
under study~\cite{fytas17b}, we start our analysis with an
alternative to this approach. In particular, we implement a three
lattice-size variant of the original quotients
method~\cite{quotients}, also known as phenomenological
renormalization~\cite{ballesteros96,amit05,nightingale76} that has
been described in detail in reference~\cite{fytas17} and already
successfully applied to the $D = 3$~\cite{fytas16} and $D =
4$~\cite{fytas16b} models. The main idea in this perspective,
given that $\alpha - 1 < 0$, is the elimination of the
non-divergent background term $E_{J,\infty}$ in
equation~(\ref{eq:E_J_scaling}) by considering three lattice sizes
in the following sequence: $(L_1, L_2, L_3) = (L, 2L, 4L)$ [see
figure~\ref{fig:2} for an instructive illustration of the
three-lattice variant of the quotients method based on the
crossings of $\xi^{\rm (con)} / L$]. Taking the quotient of the
differences at the crossings of the pairs $(L,2L)$ and $(2L,4L)$
\begin{equation}\label{eq:FSS-3L}
\hat Q_O=\frac{\left.(\overline{E}_{J,4L}-\overline{E}_{J,2L})\right|_{(\xi_{4L}/\xi_{2L})=2}}{\left(\overline{E}_{J,2L} - \overline{E}_{J,L})\right|_{(\xi_{2L}/\xi_{L})=2}}\,,
\end{equation}
one obtains the following scaling formula for the bond-energy density~\cite{fytas17}
\begin{equation}\label{eq:QO_new}
\hat Q_{\overline{E}_J}^\mathrm{(cross)}=2^{(\alpha -
1)/\nu}+\mathcal{O}(L^{-\omega}).
\end{equation}

\begin{figure}
\centerline{\includegraphics[scale=.4, angle=0]{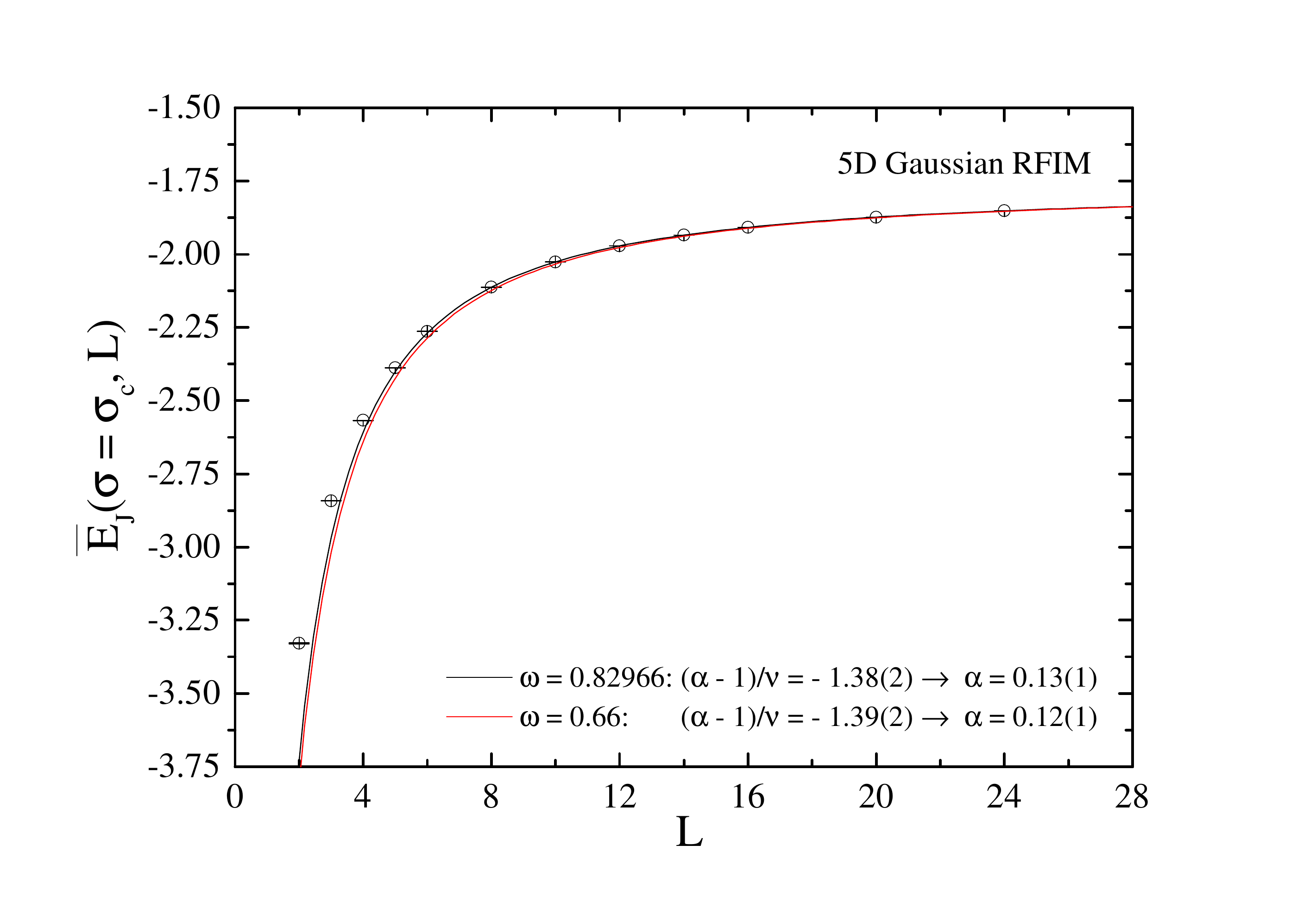}}
\caption{Finite-size scaling behavior of the bond-energy density
at the critical random-field strength $\sigma_{\rm c}(G)$ of the
5D Gaussian RFIM. The lines are fittings of the
form~(\ref{eq:E_J_scaling}) with different $\omega$ values, as
indicated in the panel.} \label{fig:4}
\end{figure}

Our results for the effective exponent ratio $(\alpha-1)/\nu$ as a
function of $L^{-\omega}$ -- where $\omega$ is set to the 3D Ising
value $0.82966$~\cite{kos16} -- are shown in figure~\ref{fig:3}.
The dashed line marks the estimate $(\alpha-1)/\nu = -1.412\;
625\;34\ldots$ of the 3D Ising universality class, where we have
used the values $\alpha = 0.110087(12)$ and $\nu =
0.629971(4)$~\cite{kos16}. A few comments are in order: (i)
Clearly, there exist large corrections to scaling for the sequence
of smaller sizes $(2,4,8)$ and $(3,6,12)$ that obscure the
application of any finite-size scaling approach. (ii) The
remaining data points [$(4,8,16)$, $(5,10,20)$, and $(6,12,24)$]
do not allow for a safe extrapolation of the ratio
$(\alpha-1)/\nu$ to $L\rightarrow \infty$, although the general
trend of the data appears to be on the right track and, in fact,
joint polynomial fits with a shared constant term do approach the
value $-1.45(6)$ but with a rather bad fitting quality. (iii)
Larger system sizes would be needed to clarify this point, but are
unfortunately out of reach with our current resources.

\begin{figure}
\centerline{\includegraphics[scale=.4, angle=0]{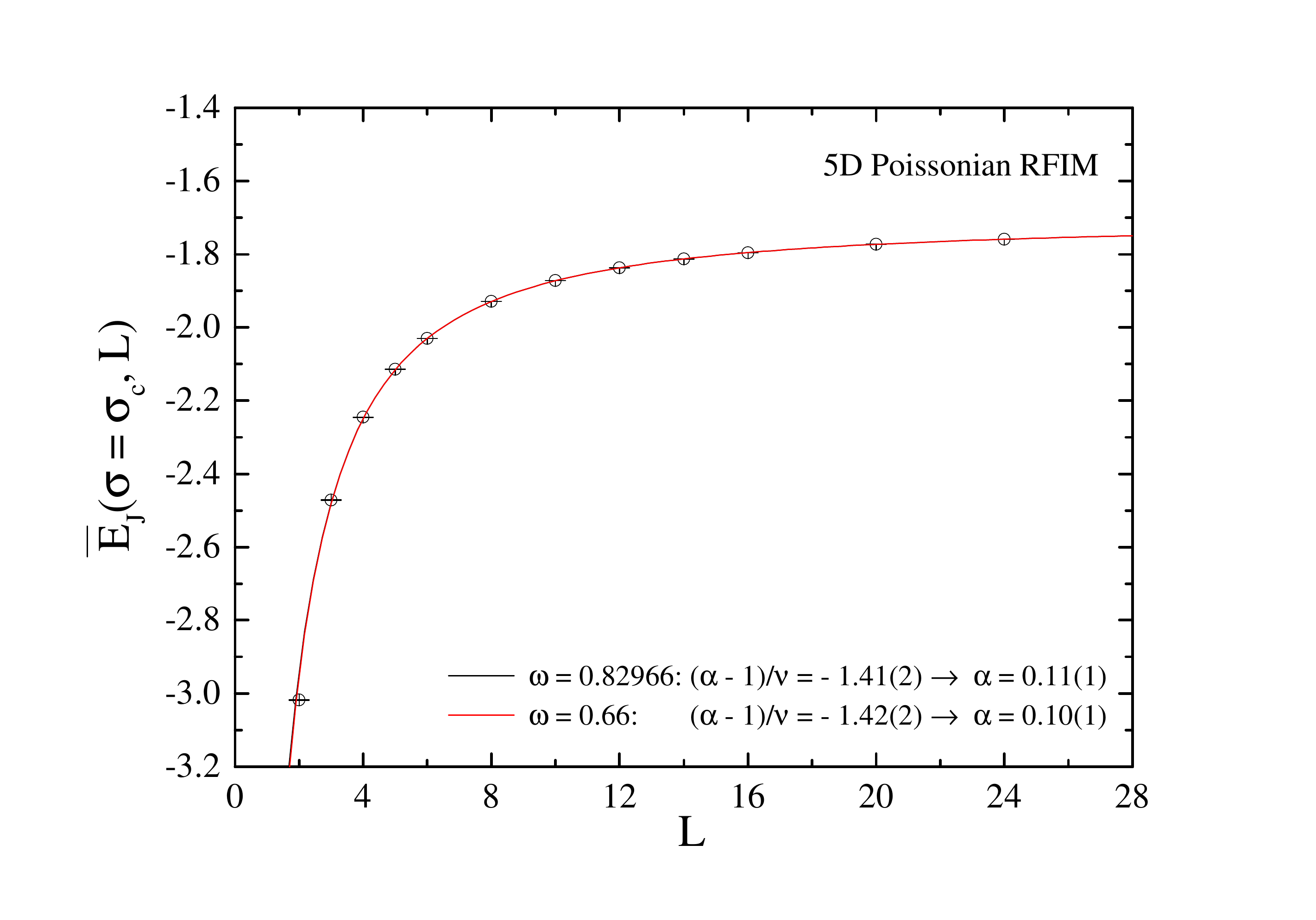}}
\caption{Finite-size scaling behavior of the bond-energy density
at the critical random-field strength $\sigma_{\rm c}(P)$ of the
5D Poissonian RFIM. The lines are fittings of the
form~(\ref{eq:E_J_scaling}) with different $\omega$ values, as
indicated in the panel.} \label{fig:5}
\end{figure}

Guided by these qualitative results of the
phenomenological-renormalization approach, we have performed, at a
second stage, additional simulations at the critical points
$\sigma_{\rm c}(G) = 6.02395$ and $\sigma_{\rm c}(P) = 5.59038$ of
the Gaussian and Poissonian models, respectively~\cite{fytas17b}.
In figures~\ref{fig:4} and \ref{fig:5} we report on the
finite-size scaling behavior of the bond-energy density at these
critical points for the whole spectrum of system sizes studied,
alongside with the resulting estimates for the ratio $(\alpha -
1)/\nu$. In both panels the solid lines are fits of the
form~(\ref{eq:E_J_scaling}), where the different colors correspond
to different fixed values of $\omega$. Black curves correspond to
the value $0.82966$ of the 3D Ising universality
class~\cite{kos16}, whereas red curves to the value $0.66$
estimated in reference~\cite{fytas17b}. The fitting quality,
measured in terms of $\chi^2$/dof, where dof measures the number
of degrees of freedom, and the minimum system size, $L_{\rm min}$,
used in the fits are as follows: $\chi^2/\rm{dof} = 1.8/3$,
$L_{\rm min} = 8$ for the Gaussian model (figure~\ref{fig:4}) and
$\chi^2/\rm{dof} = 6.1/4$, $L_{\rm min} = 6$ for the Poissonian
model (figure~\ref{fig:5}). Note that there was practically no
variation in the fitting quality moving from $\omega = 0.82966$
down to $0.66$~\cite{comment}. Using now the estimate $\nu =
0.629971(4)$ for the critical exponent of the correlation length,
simple algebra and error propagation produces values for $\alpha$
within the range $0.10 - 0.13$. Taking an average over the values
of $\alpha$ obtained from the black curves with $\omega =
0.82966$, we give our final estimate for the critical exponent
$\alpha$ to be
\begin{equation}\label{eq:alpha}
\alpha = 0.12(2).
\end{equation}
This is compatible to the value $0.12(5)$ obtained in reference~\cite{fytas17b} via the modified hyperscaling relation $\alpha = 2 - \nu(D - 2 + \bar\eta - \eta)$, where $\eta$ and $\bar\eta$ are the corresponding anomalous dimensions of the connected and disconnected correlation functions [see equation~(\ref{eq:anomalous})] and also agrees nicely with the 3D Ising universality benchmark $\alpha = 0.110087(12)$~\cite{kos16}.

\begin{figure}
\centerline{\includegraphics[scale=.4, angle=0]{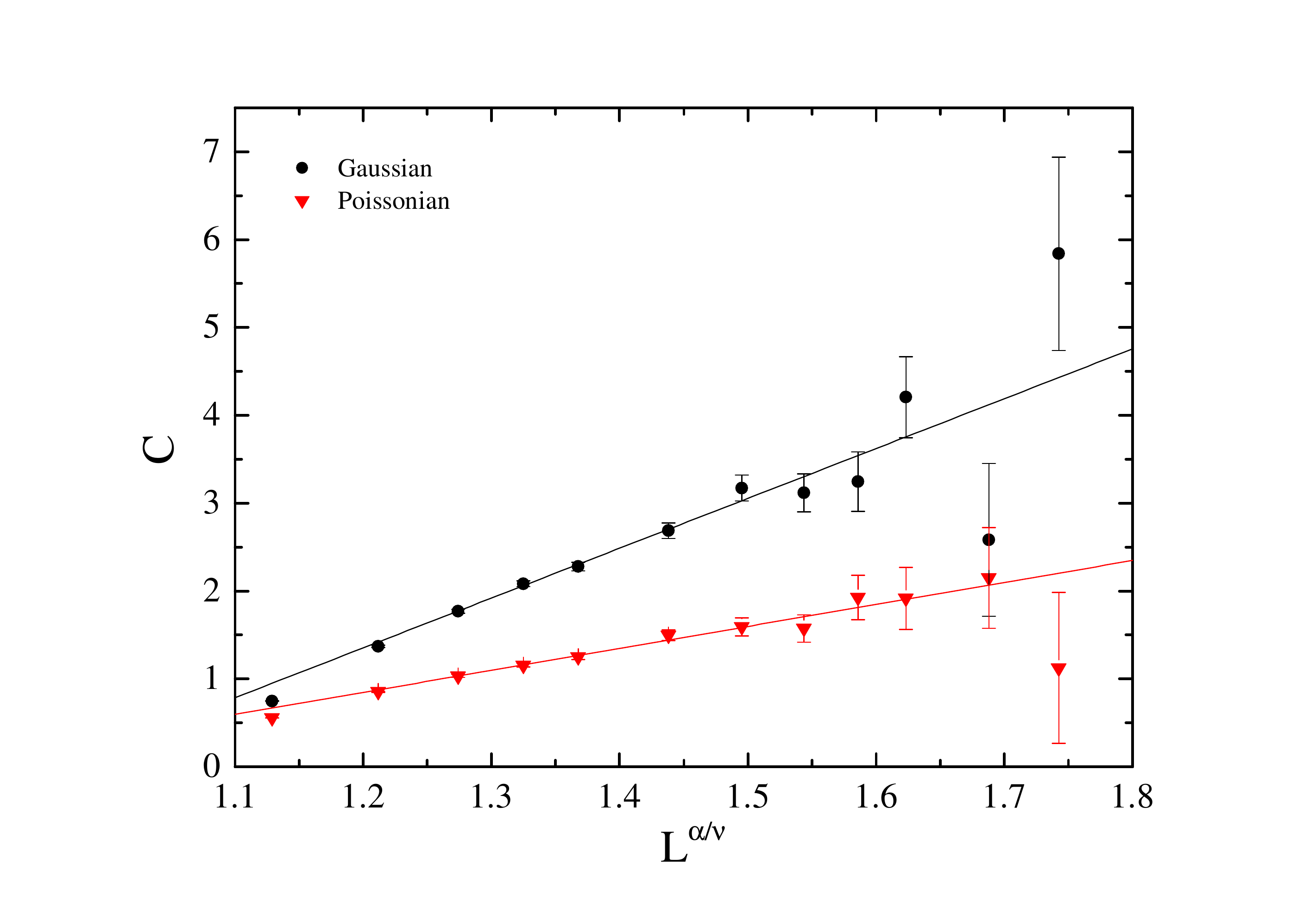}}
\caption{Scaling behavior of the specific heat $C$ for both models
considered in this work, as indicated in the panel. For a detailed
discussion on the scaling laws and the fitting tests refer to the
main text.} \label{fig:6}
\end{figure}

As an additional consistency check of our results shown in
figures~\ref{fig:4} and \ref{fig:5}, we depict in
figure~\ref{fig:6} the scaling behavior of the specific heat $C$,
obtained from the derivative of the bond-energy density with
respect to the random-field strength $\sigma$, at the critical
point. Note that the horizontal axis has been rescaled to
$L^{\alpha/\nu}$ (remember that as in the standard case $C \sim
L^{\alpha/\nu}$), and $\alpha/\nu$ has been set to the value
$0.174749\ldots$ via $\alpha = 0.110087$ and $\nu = 0.629971$ of
the 3D Ising universality class~\cite{kos16}. As expected the data
become rather noisy with increasing system size, forcing us to
exclude from our fittings the larger system sizes $L=20$ and
$L=24$, where statistical errors are larger than $30\%$. Although
we illustrate for the benefit of the reader data for the complete
spectrum of system sizes studied, the solid lines are simple
linear fits within the range $L = 4 - 16$ with a very good fitting
quality indeed: $\chi^2/\rm{dof} = 4.16/6$ and $2.03/6$ for the
Gaussian and Poissonian models, respectively.

\begin{figure}
\centerline{\includegraphics[scale=.4, angle=0]{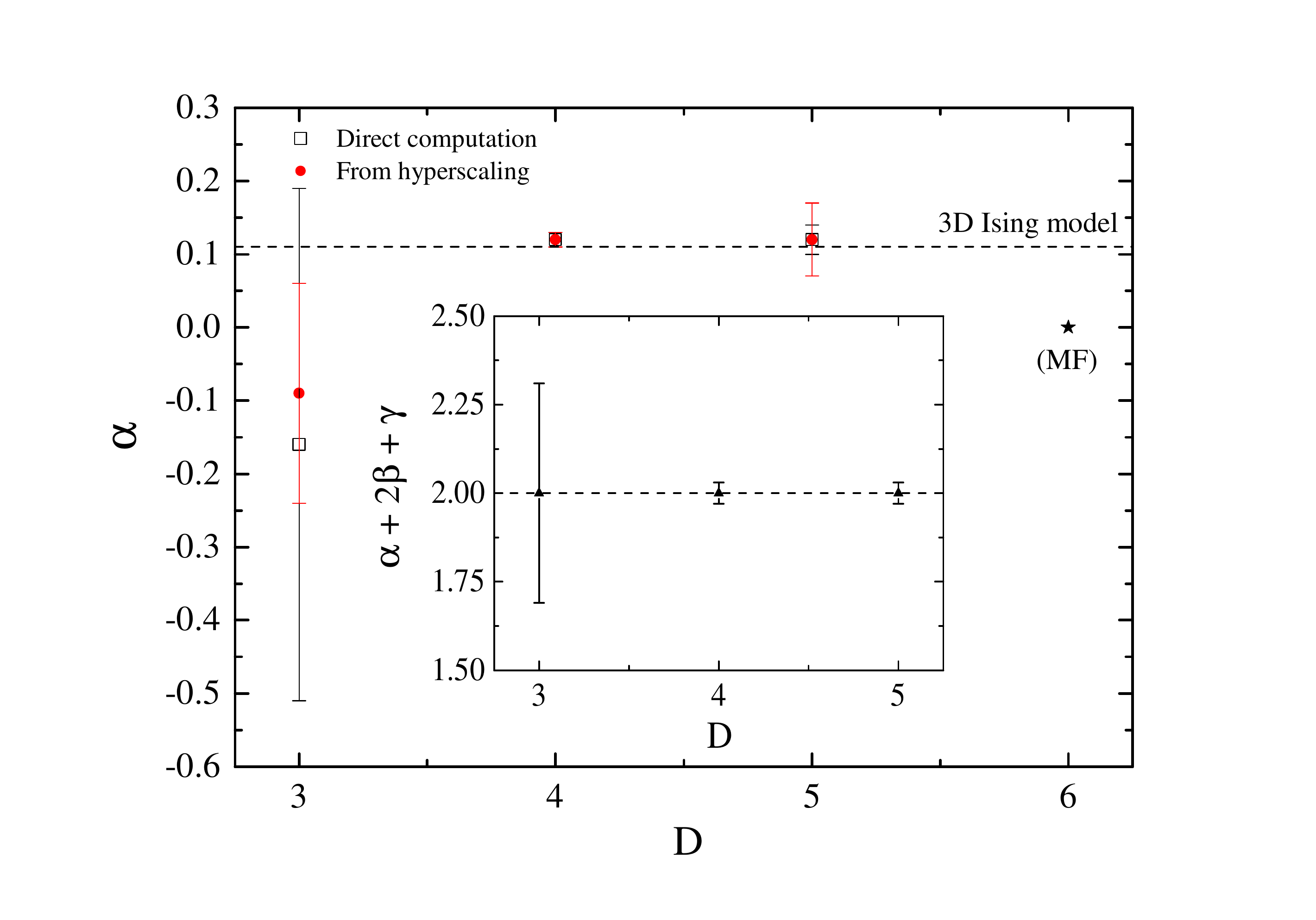}}
\caption{Critical exponent $\alpha$ of the specific heat of the RFIM as a function of the spatial dimension $D$. Two sets of data points are shown: Estimates from direct computation (open squares) and from the modified hyperscaling relation (filled circles) via previously obtained results for the exponents $\eta$, $\bar\eta$, and $\nu$. The dashed line marks the 3D Ising value $0.110087(12)$~\cite{kos16}. The filled star signals the mean-field (MF) value $\alpha = 0$, expected to hold at $D\geq 6$. {\bf Inset}: Verification of the Rushbrooke scaling relation. For the estimation of the magnetic critical exponents $\beta$ and $\gamma$ we have used the standard relations $\beta = \nu(D - 4 + \bar\eta)/2 $ and $\gamma = \nu (2-\eta)$. The dashed line is located exactly at the value $2$. Data taken from this work and from references~\cite{fytas16,fytas13,fytas16b,fytas17,fytas17b}.} \label{fig:7}
\end{figure}

To summarize, using extensive numerical simulations at zero
temperature we provided a high-precision estimate of the
specific-heat's critical exponent of the 5D RFIM. Our final result
$\alpha=0.12(2)$ is fully consistent with the estimation coming
from the modified hyperscaling relation given in
reference~\cite{fytas17b}, and also supports the recent results of
reference~\cite{fytas19} for the restoration of supersymmetry and
dimensional reduction in the RFIM at $D = 5$. We close this
contribution with figure~\ref{fig:7} and an overview of the
critical exponent $\alpha$ of the RFIM at all physically relevant
dimensions. Two sets of data points are shown, as outlined in the
caption, corroborated by a graphical validation of the Rushbrooke
relation in the corresponding inset. Whilst the collative results
of figure~\ref{fig:7} are reassuring and settle down previous
controversies in the random-field problem originating from
defective estimations of the critical exponent $\alpha$, for
reasons of clarity we sould like to point out that the large error
at $D = 3$ stems from the joint fits of $[(\alpha - 1)/\nu]^{\rm
(eff)}$ performed over several random-field distributions
(including the double Gaussian distribution) and the large scaling
corrections via $\omega(D = 3) = 0.52$~\cite{fytas16,fytas13} --
for further details and graphical explanations on this aspect we
refer the interested reader to figures 6 and 7 of
reference~\cite{fytas16}.

\section*{Acknowledgments}
We acknowledge partial financial support from Ministerio de
Econom\'ia, Industria y Competitividad (MINECO, Spain) through Grant
No. FIS2015-65078-C2 and
PGC2018-094684-B-C21, and from the European Research Council (ERC)
under the European Union's Horizon 2020 research and innovation
program (Grant No. 694925).

{}


\section*{References}
\begin{thebibliography}{}

\bibitem{imr75} Imry Y and Ma S-K 1975 \emph{Phys. Rev. Lett.} {\bf 35} 1399

\bibitem{aharony76} Aharony A, Imry Y, and Ma S-K 1976 \emph{Phys. Rev. Lett.} {\bf 37} 1364

\bibitem{young77} Young A P 1977 \emph{J. Phys. Condens. Matter} {\bf 10} L257

\bibitem{fishman79} Fishman S and Aharony A 1979 J. Phys. C: Solid State Phys. {\bf 12} L729

\bibitem{parisi79} Parisi G and Sourlas N 1979 \emph{Phys. Rev. Lett.} {\bf 43} 744

\bibitem{cardy84} Cardy J L 1984 \emph{Phys. Rev. B} {\bf 29} 505

\bibitem{imbrie84} Imbrie J Z 1984 \emph{Phys. Rev. Lett.} {\bf 53} 1747

\bibitem{villain84} Villain J 1984 \emph{Phys. Rev. Lett.} {\bf 52} 1543; Villain J 1985 \emph{J. Physique} {\bf 46} 1843

\bibitem{bray85} Bray A J and Moore M A 1985 \emph{J. Phys. C: Solid State Phys.} {\bf 18} L927

\bibitem{fisher86} Fisher D S 1986 \emph{Phys. Rev. Lett.} {\bf 56} 416

\bibitem{schwartz85} Schwartz M and Soffer A 1985 \emph{Phys. Rev. Lett.} {\bf 55} 2499 ; Schwartz M and Soffer A 1986 \emph{Phys.
Rev. B} {\bf 33} 2059 ; Schwartz M 1985 \emph{J. Phys. Condens.
Matter} {\bf 18} 135 ; Schwartz M, Gofman M, and Nattermann T
1991 \emph{Physica A} {\bf 178} 6 ; Schwartz M 1994
\emph{Europhys. Lett.} {\bf 15} 777

\bibitem{gofman93} Gofman M, Adler J, Aharony A, Harris A B, and Schwartz M 1993 \emph{Phys. Rev. Lett.} {\bf 71} 1569

\bibitem{esser97} Esser J and Nowak U 1997 \emph{Phys. Rev. B} {\bf 55} 5866

\bibitem{barber} Barber W C and Belanger D P 2001 \emph{J. Magn. Magn. Mater.} {\bf 226} 545

\bibitem{young1999} See, e.g., the articles by Belanger D P and Nattermann T 1998 {\em Spin
Glasses and Random Fields} ed A P Young (Singapore: World Scientific)

\bibitem{by1991} Belanger D P and Young A P 1991 \emph{J. Magn. Magn. Mater.} {\bf 100} 272

\bibitem{rieger1995} Rieger H 1995  \emph{Annual Reviews of Computational Physics II} ed
D Stauffer (Singapore: World Scientific) pp 295-341

\bibitem{jaccarino} Belanger D P, King A R, Jaccarino V, and Cardy J L 1983 \emph{Phys. Rev. B} {\bf 28} 2522
; Belanger D P and Slani\u{c} Z 1998 \emph{J. Magn. Magn. Mater.}
{\bf 186} 65

\bibitem{vink} Vink R L C, Binder K, and L\"{o}wen H 2006 \emph{Phys. Rev. Lett.} {\bf 97} 230603

\bibitem{berker86} Berker A N and McKay S R 1986 \emph{Phys. Rev. B} {\bf 33}, 4712

\bibitem{bricmont87} Bricmont J and Kupiainen A 1987 \emph{Phys. Rev. Lett.} {\bf 59} 1829

\bibitem{ahrens11} Ahrens B and Hartmann A K 2011 Phys. Rev. B {\bf 83} 014205

\bibitem{fytas16} Fytas N G and Mart\'{i}n-Mayor V 2016 \emph{Phys. Rev. E} {\bf 93}, 063308

\bibitem{fytas13} Fytas N G and Mart\'{i}n-Mayor V 2013 \emph{Phys. Rev. Lett.} {\bf 110} 227201

\bibitem{fytas16b} Fytas N G, Mart\'{i}n-Mayor V, Picco M, and Sourlas N 2016 \emph{Phys. Rev. Lett.}  {\bf 116} 227201

\bibitem{fytas17} Fytas N G, Mart\'{i}n-Mayor V, Picco M, and Sourlas N
\emph{J. Stat. Mech.} (2017) 033302

\bibitem{fytas17b} Fytas N G, Mart\'{i}n-Mayor V, Picco M, and Sourlas N 2017 \emph{Phys. Rev. E}  {\bf 95} 042117

\bibitem{fytas18} Fytas N G, Mart\'{i}n-Mayor V, Picco M, and Sourlas N 2018 \emph{J. Stat. Phys.}  {\bf 172} 665

\bibitem{fytas19} Fytas N G, Mart\'{i}n-Mayor V, Parisi G, Picco M, and Sourlas N 2019 \emph{Phys. Rev. Lett.}  {\bf 122} 240603

\bibitem{tissier11}  Tissier M and Tarjus G 2011 \emph{Phys. Rev. Lett.} {\bf 107} 041601

\bibitem{tissier12} Tissier M and Tarjus G 2012 \emph{Phys.  Rev.  B} {\bf 85} 104203

\bibitem{tarjus13} Tarjus G, Balog I, and Tissier M 2013 \emph{Europhys. Lett.} {\bf 103} 61001

\bibitem{hikami18} Hikami S, arXiv:1801.09052

\bibitem{middleton1} Middleton A A and Fisher D S 2002 \emph{Phys. Rev. B} {\bf 65} 134411

\bibitem{hartmann01} Hartmann A K and Young A P 2001 \emph{Phys. Rev. B} {\bf 64} 214419

\bibitem{theodorakis} Theodorakis P E, Georgiou I, and Fytas N G 2013 \emph{Phys. Rev. E} {\bf 87} 032119

\bibitem{fytasEPJB} Fytas N G, Theodorakis P E, Georgiou I, and Lelidis I 2013 \emph{Eur. Phys. J. B} {\bf 86} 268

\bibitem{nowak} Nowak U, Usadel K D, and Esser J 1998 \emph{Physica A} {\bf 250}, 1

\bibitem{kos16} Kos F, Poland D, Simmons-Duffin D, and Vichi A 2016 \emph{J. High
Energy Phys.} {\bf 08} 036

\bibitem{ogielski85} Ogielski A T 1986 {\emph Phys. Rev. Lett.} {\bf 57} 1251

\bibitem{hartmann95} Hartmann A K and Usadel K D 1995 \emph{Physica A} {\bf 214} 141;
Hartmann A K 1998 \emph{Physica A} {\bf 248} 1

\bibitem{bastea98} Bastea S and Duxbury P M 1998 \emph{Phys. Rev. E} {\bf 58} 4261 ; Bastea S 1998 \emph{Phys. Rev. E} {\bf 58}
7978 ; Bastea S and Duxbury P M 1999 \emph{Phys. Rev. E} {\bf 60}
4941

\bibitem{hartmann99} Hartmann A K and Nowak U 1999 \emph{Eur. Phys. J. B} {\bf 7} 105

\bibitem{hartmann02} Hartmann A K 2002 \emph{Phys. Rev. B} {\bf 65} 174427

\bibitem{seppala} Sepp\"{a}l\"{a} E T and Alava M J 2001 \emph{Phys. Rev. E} {\bf 63} 066109 ; Sepp\"{a}l\"{a} E T,
Alava M J, and Duxbury P M 2001 \emph{Phys. Rev. E} {\bf 63}
066110 ; Sepp\"{a}l\"{a} E T, Pulkkinen A M, and Alava M J 2002
\emph{Phys. Rev. B} {\bf 66} 144403

\bibitem{middleton2} Middleton A A 2002 \emph{Phys. Rev. Lett.} {\bf 88} 017202

\bibitem{middleton3} Middleton A A arXiv:cond-mat/0208182 ; Meinke J H and Middleton A A
arXiv:cond-mat/0502471 ; Hambrick D C, Meinke J H, and Middleton A
A arXiv:cond-mat/0501269

\bibitem{machta03} Dukovski I and Machta J 2003 \emph{Phys. Rev. B} {\bf 67} 014413

\bibitem{alava} Alava M J, Duxbury P M, Moukarzel C F, and Rieger H 2001 \emph{Phase Transitions and Critical Phenomena},
vol 18, ed C Domb and J L Lebowitz (San Diego, CA: Academic)

\bibitem{zumsande} Zumsande M, Alava M J, and Hartmann A K 2008 \emph{J. Stat. Mech.: Theory Exp.} P02012

\bibitem{puri11} Shrivastav G P, Krishnamoorthy S, Banerjee V, and Puri S 2011 \emph{Europhys. Lett.} {\bf 96} 36003

\bibitem{weigel} Stevenson J D and Weigel M 2011 \emph{Europhys. Lett.} {\bf 95} 40001

\bibitem{fytas12} Fytas N G, Theodorakis P E, and Georgiou I 2012 \emph{Eur. Phys. J. B} {\bf 85} 349

\bibitem{hartmannbook1} Hartmann A K and Rieger H 2004 \emph{Optimization Algorithms in Physics}  (Berlin: Wiley)

\bibitem{hartmannbook2} Hartmann A K and Weigt M 2005 \emph{Phase Transitions in Combinatorial Optimization Problems} (Berlin: Wiley)

\bibitem{angles} Angl\`{e}s d'Auriac J-C, Preissmann M, and Rammal R 1985 \emph{J. Phys. Lett.} {\bf 46} L173

\bibitem{cormen} Cormen T H, Leiserson C E, and Rivest R L 1990 \emph{Introduction To Algorithms}, (Cambridge: MA MIT Press)

\bibitem{papadimitriou} Papadimitriou C H 1994 \emph{Computational Complexity}  (Reading: Addison-Wesley)

\bibitem{tarjan} Goldberg A V and Tarjan R E 1988 \emph{J. Assoc. Comput. Mach.} {\bf 35}
921 ; Cherkassky B V and Goldberg A V 1997 Algorithmica {\bf 19}
390


\bibitem{holm97} Holm C and Janke W 1997 \emph{Phys. Rev. Lett.} {\bf 78} 2265

\bibitem{quotients} The general approach in the quotients method is to compare
observables computed in pair of lattices $(L,2L)$. We start
imposing scale-invariance by seeking the $L$-dependent critical
point: the value of $\sigma$ such that $\xi_{2L}/\xi_L=2$, i.e.,
the crossing point for $\xi_L/L$ [see also figure~\ref{fig:2} for the case where $\xi = \xi^{\rm (con)}$]. For dimensionful quantities $O$, scaling in the thermodynamic limit as $\xi^{x_O/\nu}$, we consider the quotient
$Q_O=O_{2L}/O_L$ at the crossing. Thus, we have $Q_O^\mathrm{(cross)}=2^{x_O/\nu}+\mathcal{O}(L^{-\omega})$, where $x_O/\nu$, and the scaling-corrections exponent $\omega$ are universal

\bibitem{ballesteros96} Ballesteros H G, Fern\'{a}ndez L A, Mart\'{i}n-Mayor V, and
Mu\~{n}oz Sudupe A 1996 \emph{Phys. Lett. B} {\bf 378} 207

\bibitem{amit05} Amit D J and Mart\'{i}n-Mayor V 2005 \emph{ Field Theory, the
Renormalization Group and Critical Phenomena}, 3rd edn (Singapore: World
Scientific)

\bibitem{nightingale76} Nightingale M 1976 \emph{Physica A} {\bf 83} 561

\bibitem{comment} See also the relevant statistical tests with respect to the value of $\omega$ for the other thermodynamic observables in reference~\cite{fytas17b}

\end{thebibliography}
\end{document}